\documentclass[conference]{IEEEtran}
\usepackage{amssymb}
\usepackage{amsmath}
\usepackage{cite}
\usepackage{url}
\usepackage{xcolor}
\usepackage{cite,graphicx,amsmath,amssymb}
\usepackage{subfigure}
\usepackage{citesort}
\usepackage{fancyhdr}
\usepackage{mdwmath}
\usepackage{mdwtab}
\usepackage{caption}
\usepackage{amsthm}
\usepackage{algorithm}
\usepackage{algorithmic}
\usepackage{makecell}
\usepackage{diagbox}
\allowdisplaybreaks
\usepackage[
top    = 1.70cm,
bottom = 1.05in,
left   = 0.63 in,
right  = 0.63 in]{geometry}

\newtheorem{remark}{Remark}
\newtheorem{theorem}{Theorem}

\newtheorem{lemma}{Lemma}

\newtheorem{corollary}{Corollary}

\newcommand{\bm}[1]{\mbox{\boldmath{$#1$}}}
\begin{document}
\title{Joint Radar and Multicast-Unicast Communication: A NOMA Aided Framework}

\author{\IEEEauthorblockN{Xidong Mu\IEEEauthorrefmark{1}\IEEEauthorrefmark{4},Yuanwei Liu\IEEEauthorrefmark{2},
Li Guo\IEEEauthorrefmark{1}\IEEEauthorrefmark{4}, Jiaru Lin,\IEEEauthorrefmark{1}\IEEEauthorrefmark{4} and Lajos~Hanzo\IEEEauthorrefmark{3}}
\IEEEauthorblockA{\IEEEauthorrefmark{1}Key Lab of Universal Wireless Communications, Ministry of Education,\\Beijing University of Posts and Telecommunications, Beijing, China. \\\IEEEauthorrefmark{4}School of Artificial Intelligence, Beijing University of Posts and Telecommunications, Beijing, China.\\ \IEEEauthorrefmark{2}Queen Mary University of London, London, U.K.\\\IEEEauthorrefmark{3}School of Electronics and Computer Science, the University of Southampton, Southampton, U.K.\\E-mail:\{muxidong, guoli, jrlin\}@bupt.edu.cn, yuanwei.liu@qmul.ac.uk, lh@ecs.soton.ac.uk}}

\maketitle
\vspace{-1.3cm}
\begin{abstract}
The novel concept of non-orthogonal multiple access (NOMA) aided joint radar and multicast-unicast communication (Rad-MU-Com) is investigated. Employing the same spectrum resource, a multi-input-multi-output (MIMO) dual-functional radar-communication (DFRC) base station detects the radar-centric user (R-user), while transmitting mixed multicast-unicast messages both to the R-user and to the communication-centric user (C-user). In particular, the multicast information is intended for both the R- and C-users, whereas the unicast information is only intended for the C-user. More explicitly, NOMA is employed to facilitate this \emph{double spectrum sharing}, where the multicast and unicast signals are superimposed in the power domain and the superimposed communication signals are also exploited as radar probing waveforms. A \emph{beamformer-based} NOMA-aided joint Rad-MU-Com framework is proposed for the system having a single R-user and a single C-user. Based on this framework, the unicast rate maximization problem is formulated by optimizing the beamformers employed, while satisfying the rate requirement of multicast and the predefined accuracy of the radar beam pattern. The resultant non-convex optimization problem is solved by a penalty-based iterative algorithm to find a high-quality near-optimal solution. Finally, our numerical results reveal that significant performance gains can be achieved by the proposed scheme over the benchmark schemes.
\end{abstract}
\section{Introduction}
Given the rapid development of cost-efficient electronic technologies, the number of connected devices (e.g., smart phones and Internet-of-Things (IoT) nodes) in the wireless networks escalates. It is forecast that the global mobile data traffic in 2022 will be seven times of that in 2017~\cite{Cisco}. Moreover, new attractive applications (e.g., virtual reality (VR), augmented reality (AR), and ultra-high definition (UHD) video streaming) have emerged, which significantly improve the user-experience, but exacerbate the spectral congestion. As a remedy, a promising solution is to harness spectrum sharing between radar and communication systems~\cite{survey}.\\
\indent Radar (which is short for ``radio detection and ranging'') was originally proposed for military applications in the 1930s, and has rapidly developed in the past decades for both civilian and military applications~\cite{radar}. In contrast to wireless communications, where the radio waves convey information bits, radar employs radio waves to determine the target's characteristics (e.g., location, velocity, shape, etc.) by first transmitting probing signals and then analyzing the received echoes reflected by the target. By carrying out spectrum sharing between radar and communication, on the one hand, communication systems are allowed to glean additional spectrum resources, which are occupied by radar systems (e.g., the S-band (2-4 GHz) and C-band (4-8 GHz))~\cite{survey,survey2}. On the other hand, the integration of radar and communication would support promising but challenging near-future applications, such as autonomous vehicles (AVs) and smart homes.\\
\indent In recent years, dual-function radar communication (DFRC)~\cite{DFRC} has been proposed as a promising technology, where the functions of radar and communication are facilitated using a joint platform, thus reducing the resultant hardware cost. Given these advantages, DRFC has become a focal point of the CRSS research field. For instance, Liu {\em et al.}~\cite{protocol} proposed a pair of sophisticated strategies for implementing DFRC, namely a separated and a shared deployment, with the aim of constructing a high-quality radar beam pattern, while satisfying the communication requirements. As a further development, based on the separated and shared deployment, Dong {\em et al.}~\cite{low} and Liu {\em et al.}~\cite{radio} conceived low-complexity BF design algorithms, respectively. As an innovative contribution, Liu {\em et al.}~\cite{optimal} studied the optimal waveform design of DFRC under the shared deployment paradigm, where branch-and-bound based algorithms were developed. Su {\em et al.}~\cite{security} studied the secure transmission for DFRC, where the radar target was treated as an eavesdropper and artificial noise was employed for degrading its received SINR, while satisfying the communication requirements of the legitimate users. Note that the aforementioned research contributions on DFRC only considered unicast communications and assumed that the radar target does not communicate, it merely has to be detected. Given the diverse future applications of wireless networks, more sophisticated CRSS schemes have to be conceived for supporting mixed multicast-unicast communication and simultaneously communicating with and detecting the radar target user, which, to the best of our knowledge, has not been investigated. This provides the main motivation of this work.\\
\indent In this paper, we investigate a non-orthogonal multiple access (NOMA)-aided joint radar and multicast-unicast communication (Rad-MU-Com) system, which consists of two types of users, namely the radar-centric user (R-user) and the communication-centric user (C-user). By employing power-domain NOMA, a \emph{double spectrum sharing} operation is facilitated, where the MIMO DFRC base station (BS) transmits the mixed multicast and unicast messages to the R- and C-users, while detecting the R-user target using the transmitted superimposed communication signals. In particular, we propose a beamformer-based NOMA (BB NOMA)-aided joint Rad-MU-Com framework, where the multicast and unicast messages are transmitted via different BFs. Based on this, we formulate a BF design problem for the maximization of the unicast rate, subject to both the multicast rate requirement and to the radar beam pattern accuracy achieved. To solve the resultant non-convex problem, we develop an efficient penalty-based iterative algorithm for finding a stationary point of the original optimization problem. Our numerical results show that the proposed BB NOMA-aided joint Rad-MU-Com scheme achieves a higher unicast performance than the benchmark schemes relying on conventional transmission strategies. Furthermore, the performance gain becomes more significant, when the constraint on the radar beam pattern is more relaxed.\\
\indent \emph{Notations:} Scalars, vectors, and matrices are denoted by lower-case, bold-face lower-case, and bold-face upper-case letters, respectively; ${\mathbb{C}^{N \times 1}}$ denotes the space of $N \times 1$ complex-valued vectors; ${{\mathbf{a}}^H}$ and $\left\| {\mathbf{a}} \right\|$ represent the conjugate transpose of vector ${\mathbf{a}}$; ${\mathcal{CN}}\left( {\mu,\sigma ^2} \right)$ denotes the distribution of a circularly symmetric complex Gaussian (CSCG) random variable with mean $\mu $ and variance ${\sigma ^2}$; ${{\mathbf{1}}}$ stands for the all-one vector; ${\rm {Rank}}\left( \mathbf{A} \right)$ and ${\rm {Tr}}\left( \mathbf{A} \right)$ denote the rank and the trace of matrix $\mathbf{A}$, respectively; ${\rm {Diag}}\left( \mathbf{A} \right)$ represents a vector whose elements are extracted from the main diagonal elements of matrix $\mathbf{A}$; ${{\mathbf{A}}} \succeq 0$ indicates that $\mathbf{A}$ is a positive semidefinite matrix; ${\mathbb{H}^{N}}$ denotes the set of all $N$-dimensional complex Hermitian matrices. ${\left\| {\mathbf{A}} \right\|_*}$, ${\left\| {\mathbf{A}} \right\|_2}$, and ${\left\| {\mathbf{A}} \right\|_F}$ are the nuclear norm, spectral norm, and Frobenius norm of matrix $\mathbf{A}$, respectively.
\section{System Model and Problem Formulation}
\subsection{System Model}
\begin{figure}[!ht]
  \centering
  \includegraphics[width=3.5in]{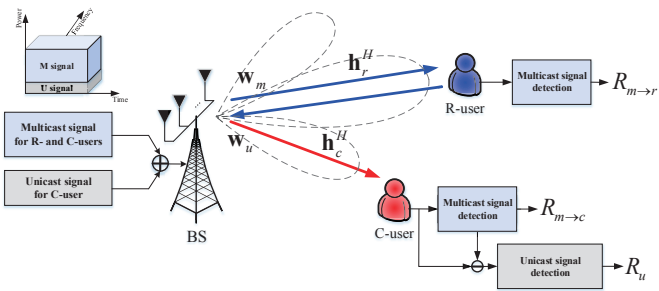}\\
  \caption{Illustration of the BB NOMA-aided joint Rad-MU-Com system.}\label{beamformer}
\end{figure}
\vspace{-0.4cm}
As illustrated in Fig. \ref{beamformer}, a MIMO DFRC system is considered, which consists of a single $N$-antenna DFRC BS, a single-antenna R-user, and a single-antenna C-user. In contract to existing work~\cite{protocol,low,radio,optimal,security}, where the DFRC BS detects the R-user located within the angles of interest while only communicating with the C-user, we consider mixed multicast-unicast transmission. To be more specific, two different types of messages have to be sent by the BS, one for both the R- and C-users, namely the multicast signal, while the unicast signal is only intended for the C-user. It is worth mentioning that this mixed multicast-unicast transmission represents the evolution of DFRC from \emph{isolation} to \emph{integration}. For instance, the multicast signal can be employed for broadcasting group-oriented system configurations and automatic software updates, which are requested both by the R- and C-users. By contrast, the unicast signal consists of personalized voice and video traffic intended for the C-user, which is not relevant for the R-user. To support this novel concept for DFRC, we propose a BB NOMA-aided joint Rad-MU-Com framework. In the following, the communication model and radar model of the proposed system will be introduced.
\subsubsection{BB NOMA-aided MU-Communication Model} For supporting our mixed multicast-unicast based MIMO DFRC system, the BB NOMA scheme of~\cite{multiple} is employed. Explicitly, the BS employs different BFs for transmitting the multicast signal intended for both R- and C-users and for the unicast signal only intended for the C-user, where the pair of signals are multiplexed in the power domain. Let $s_{m}\left[ n \right]$ and $s_{u}\left[ n \right]$ denote the multicast signal and the unicast signal at the time index $n$, respectively. Therefore, the corresponding transmitted superimposed signal at the $n$th time index is given by
\vspace{-0.2cm}
\begin{align}\label{transmit signal}
{\mathbf{x}}_1\left[ n \right] = {{\mathbf{w}}_{m}}s_{m}\left[ n \right] + {{\mathbf{w}}_u}s_{u}\left[ n \right],
\end{align}
\vspace{-0.6cm}

\noindent where ${{\mathbf{w}}_{m}} \in {{\mathbb{C}}^{N \times 1}}$ and ${{\mathbf{w}}_u}\in {{\mathbb{C}}^{N \times 1}}$ represents the BFs designed for transmitting the multicast and unicast information symbols, respectively. Without loss of generality, we assume that the multicast and unicast signals are statistically independent of each other and we have ${s_{m}}\left[ n \right] \sim {\mathcal{C}\mathcal{N}}\left( {0,1} \right)$ and ${s_u}\left[ n \right] \sim {\mathcal{C}\mathcal{N}}\left( {0,1} \right)$. Let ${{\mathbf{h}_r^H}}\in {{\mathbb{C}}^{1 \times N}}$ and ${{\mathbf{h}_c^H}}\in {{\mathbb{C}}^{1 \times N}}$ denote the BS-R-user channel and the BS-C-user channel, respectively, which are assumed to be perfectly estimated. For the R- and C-users, the signal received at time index $n$ can be respectively expressed as
\vspace{-0.2cm}
\begin{align}\label{received signal T}
y_r\left[ n \right] = {\mathbf{h}_r^H}\left( {{{\mathbf{w}}_{m}}s_{m}\left[ n \right] + {{\mathbf{w}}_{u}}s_{u}\left[ n \right]} \right) + z_r\left[ n \right],
\end{align}
\vspace{-0.7cm}
\begin{align}\label{received signal C}
y_c\left[ n \right] = {\mathbf{h}_c^H}\left( {{{\mathbf{w}}_{m}}s_{m}\left[ n \right] + {{\mathbf{w}}_{u}}s_{u}\left[ n \right]} \right) + z_c\left[ n \right],
\end{align}
\vspace{-0.6cm}

\noindent where $z_r\left[ n \right]\sim {\mathcal{CN}}\left( {0,\sigma_r ^2} \right)$ and $z_c\left[ n \right]\sim {\mathcal{CN}}\left( {0,\sigma_c ^2} \right)$ denote the additive white Gaussian noise (AWGN) of the R- and C-users at the time index $n$, respectively. Similar to the ``strong'' user in the conventional twin-user downlink NOMA transmission [26], the potentially stronger multicast signal is detected first at the C-user. Then, the remodulated multicast signal is subtracted from the composite received signal, automatically leaving the interference-free decontaminated unicast signal behind. As a result, the achievable rate for the multicast message at the C-user is given by
\vspace{-0.4cm}
\begin{align}\label{CT rate C}
{R_{m \to c}} = {\log _2}\left( {1 + \frac{{{{\left| {{\mathbf{h}}_c^H{{\mathbf{w}}_{m}}} \right|}^2}}}{{{{\left| {{\mathbf{h}}_c^H{{\mathbf{w}}_u}} \right|}^2} + \sigma _c^2}}} \right).
\end{align}
\vspace{-0.4cm}

\noindent After subtracting the remodulated multicast signal from the composite received signal by SIC, the achievable rate for the unicast signal at the C-user is given by
\begin{align}\label{C rate C}
{R_{u}} = {\log _2}\left( {1 + \frac{{{{\left| {{\mathbf{h}}_c^H{{\mathbf{w}}_u}} \right|}^2}}}{{\sigma _c^2}}} \right).
\end{align}
\vspace{-0.4cm}

\noindent Similar to the ``weak'' user in the conventional twin-user downlink NOMA transmission~\cite{Liu2017}, the R-user directly detects the multicast signal by treating the unicast signal as noise. Therefore, the achievable rate for the multicast message can be expressed as
\vspace{-0.1cm}
\begin{align}\label{CT rate T}
{R_{m \to r}} = {\log _2}\left( {1 + \frac{{{{\left| {{\mathbf{h}}_r^H{{\mathbf{w}}_{m}}} \right|}^2}}}{{{{\left| {{\mathbf{h}}_r^H{{\mathbf{w}}_u}} \right|}^2} + \sigma _r^2}}} \right).
\end{align}
\vspace{-0.3cm}

\noindent The rate of the multicast signal is limited by the lower one of the pair of communication rates, which is given by
\vspace{-0.2cm}
\begin{align}\label{CT rate}
{R_{m}} = \min \left\{ {{R_{m \to c}},{R_{m \to r}}} \right\}.
\end{align}
\vspace{-0.6cm}
\subsubsection{BB NOMA-aided Radar Detection Model} According to \cite{protocol}, the above superimposed communication signals can also be exploited as radar probing waveforms, i.e., each transmitted information symbol can also be considered as a snapshot of a radar pulse. Therefore, the radar beam pattern design is equivalent to the design of the covariance matrix of the transmitted signal, ${\mathbf{x}}_1\left[ n \right]$, which is given by
\vspace{-0.1cm}
\begin{align}\label{covariance matrix}
{\mathbf{R}_1} = {\mathbb{E}}\left[ {{\mathbf{x}}_1\left[ n \right]{{\mathbf{x}}_1^H}\left[ n \right]} \right] = {{\mathbf{w}}_{m}}{\mathbf{w}}_{m}^H + {{\mathbf{w}}_u}{\mathbf{w}}_u^H.
\end{align}
\vspace{-0.6cm}

\noindent Then, the transmit beam pattern used for radar detection can be expressed as
\vspace{-0.1cm}
\begin{align}\label{beam pattern}
P\left( \theta  \right) = {{\bm{\alpha}} ^H}\left( \theta  \right){\mathbf{R}_1}{\bm{\alpha}} \left( \theta  \right),
\end{align}
\vspace{-0.6cm}

\noindent where ${\bm{\alpha}} \left( \theta  \right) = {\left[ {1,{e^{j\frac{{2\pi d}}{\lambda }\sin \theta }}, \ldots ,{e^{j\frac{{2\pi d}}{\lambda }\left( {N - 1} \right)\sin \theta }}} \right]^T}\in {{\mathbb{C}}^{N \times 1}}$ denotes the steering vector of the transmit antenna array, $\theta $ is the detection angle, $d$ represents the antenna spacing, and $\lambda$ is the carrier wavelength.
\vspace{-0.2cm}
\begin{remark}\label{benefits}
\emph{The main benefits of the proposed BB NOMA-aided joint Rad-MU-Com framework can be summarized as follows. Firstly, the employment of NOMA ensures the quality of the unicast transmission (which is usually data-hungry) for the C-user, since the inter-signal interference is canceled by SIC, see \eqref{C rate C}. Secondly, despite the presence of interference, the rate-requirement of both the R- and C-users can be readily guaranteed as a benefit of the power sharing provided by NOMA, see \eqref{CT rate C} and \eqref{CT rate T}. Thirdly, the different BFs used in our BB NOMA structure provide additional degrees-of-freedom (DoFs) for our radar beam pattern design, see \eqref{covariance matrix}. Last but not least, NOMA facilitates \emph{double spectrum sharing} between both the multicast and unicast as well as between radar and communication systems, thus further enhancing the SE.}
\end{remark}
\vspace{-0.4cm}
\begin{remark}\label{bechmark schems}
\emph{The joint Rad-MU-Com concept may also be facilitated by existing conventional transmission schemes. For example, the multicast and unicast signals can be successively transmitted via different time slots while detecting the R-user target, namely by a time division multiple access (TDMA) based Rad-MU-Com system. Moreover, the multicast and unicast can be transmitted via conventional BFs dispensing with SIC~\cite{protocol,low,radio,optimal,security} while detecting the R-user target, namely by a CBF-No-SIC based Rad-MU-Com system. These options will serve as the benchmark schemes in our performance comparisons of Section IV.}
\end{remark}
\vspace{-0.4cm}
\subsection{Problem Formulation}
\vspace{-0.1cm}
Before formulating the associated optimization problem, we first introduce the concept of the ideal radar beam pattern, which can be obtained by solving the following least-squares problem~\cite{protocol,idea_pattern}:
\vspace{-0.3cm}
\begin{subequations}\label{ideal beam pattern design}
\begin{align}
\mathop {\min }\limits_{\delta ,{{\mathbf{R}}_0}} &\;\Delta \left( {{{\mathbf{R}}_0},\delta } \right) \triangleq {\sum\nolimits_{m = 1}^M {\left| {\delta {P^*}\left( {{\theta _m}} \right) - {{\bm{\alpha}} ^H}\left( {{\theta _m}} \right){{\mathbf{R}}_0}{\bm{\alpha}} \left( {{\theta _m}} \right)} \right|} ^2}  \\
\label{average power}{\rm{s.t.}}\;\;&{\rm{Diag}}\left( {{{\mathbf{R}}_0}} \right) = \frac{{{P_{\max }}{\mathbf{1}}}}{N},\\
\label{SEMI}&{\mathbf{R}}_0 \succeq 0,{\mathbf{R}}_0 \in {{\mathbb{H}}^N},\\
\label{scale}&\delta \ge 0,
\end{align}
\end{subequations}
\vspace{-0.7cm}

\noindent where $\left\{ {{\theta _m}} \right\}_{m = 1}^M$ denotes an angular grid covering the detector's angular range in $\left[ { - \frac{\pi }{2},\frac{\pi }{2}} \right]$, ${\bm{\alpha}} \left( {{\theta _m}}\right)$ is the corresponding steering vector, ${P^*}\left( {{\theta _m}}\right)$ represents the desired ideal beam pattern gain at ${{\theta _m}}$, $\delta$ is a scaling factor, $P_{\max}$ is the maximum transmit power budget at the MIMO DFRC BS, and ${\mathbf{R}_0}$ is the waveform's covariance matrix, when only the MIMO radar is considered. It can be readily verified that the ideal radar beam pattern design problem of \eqref{ideal beam pattern design} is convex, which can be efficiently solved. Let ${\mathbf{R}_0^*}$ and $\delta^*$ denote the optimal solutions of \eqref{ideal beam pattern design}. The corresponding objective function value $\Delta \left( {{\mathbf{R}}_0^*,{\delta ^*}} \right)$ characterizes the minimum beam pattern error between the desired ideal beam pattern gain and the radar-only beam pattern gain. However, for supporting both the communication and radar functions in the MIMO DFRC system considered, a radar performance loss will occur. In the following, $\Delta \left( {{\mathbf{R}}_0^*,{\delta ^*}} \right)$ will be used as a performance benchmark for quantifying the radar performance loss in the joint Rad-MU-Com system design.\\
\indent Given our BB NOMA-aided joint Rad-MU-Com framework and the radar performance benchmark $\Delta \left( {{\mathbf{R}}_0^*,{\delta ^*}} \right)$, we aim for maximizing the unicast rate achieved at the C-user, while satisfying the minimum rate requirement of multicast communication at both the R- and C-users as well as achieving the desired beam pattern for radar detection. The resultant optimization problem can be formulated as follows:
\vspace{-0.3cm}
\begin{subequations}\label{beamformer based 0}
\begin{align}
&\mathop {\max }\limits_{{{\mathbf{w}}_{m}},{{\mathbf{w}}_u},{{\mathbf{R}}_1}} \;{R_{u}} \\
\label{CT QoS 2}{\rm{s.t.}}\;\;&{R_{m}} \ge {{\overline R}_{m}},\\
\label{radar QoS}&\frac{{\Delta \left( {{{\mathbf{R}}_1},{\delta ^*}} \right) - \Delta \left( {{\mathbf{R}}_0^*,{\delta ^*}} \right)}}{{\Delta \left( {{\mathbf{R}}_0^*,{\delta ^*}} \right)}} \le {{\overline \gamma }_b},\\
\label{power Communication}&{\rm{Tr}}\left( {{{\mathbf{R}}_1}} \right) = {P_{\max }},
\end{align}
\end{subequations}
\vspace{-0.6cm}

\noindent where $\overline R_{m}$ represents the minimum rate requirement of multicast, and $\overline \gamma_b$ is the maximum tolerable radar beam pattern mismatch ratio between the beam pattern error achieved in the joint Rad-MU-Com system (i.e., $\Delta \left( {{\mathbf{R}}_1,{\delta ^*}} \right)$) and the minimum one (i.e., $\Delta \left( {{\mathbf{R}}_0^*,{\delta ^*}} \right)$) obtained by the radar-only system.
\vspace{-0.1cm}
\section{Proposed Solution}
\vspace{-0.1cm}
The main challenge in solving problem \eqref{beamformer based 0} is that the objective function and the left-hand-side (LHS) is not concave with respect to the optimization variables. To address this issue, we define ${{\mathbf{W}}_{m}} = {{\mathbf{w}}_{m}}{\mathbf{w}}_{m}^H$ and ${{\mathbf{W}}_{u}} = {{\mathbf{w}}_{u}}{\mathbf{w}}_{u}^H$, which satisfy that ${{\mathbf{W}}_{m}}\succeq 0$, ${{\mathbf{W}}_{u}}\succeq 0$, ${\rm{Rank}}\left( {{{\mathbf{W}}_{m}}} \right)=1$, and ${\rm{Rank}}\left( {{{\mathbf{W}}_u}} \right) = 1$. Then, problem \eqref{beamformer based 0} can be reformulated as follows
\vspace{-0.2cm}
\begin{subequations}\label{beamformer based 1}
\begin{align}
&\mathop {\max }\limits_{{{\mathbf{W}}_{m}},{{\mathbf{W}}_u},{\mathbf{R}_1}} \;{\log _2}\left( {1 + \frac{{{\rm{Tr}}\left( {{{\mathbf{H}}_c}{{\mathbf{W}}_u}} \right)}}{{\sigma _c^2}}} \right)\\
\label{CT C QoS}{\rm{s.t.}}\;\;&{\rm{Tr}}\left( {{{\mathbf{H}}_c}{{\mathbf{W}}_{m}}} \right) - {\overline \gamma  _{m}}{\rm{Tr}}\left( {{{\mathbf{H}}_c}{{\mathbf{W}}_u}} \right) - {\overline \gamma  _{m}}\sigma _c^2 \ge 0,\\
\label{CT T QoS}&{\rm{Tr}}\left( {{{\mathbf{H}}_r}{{\mathbf{W}}_{m}}} \right) - {\overline \gamma  _{m}}{\rm{Tr}}\left( {{{\mathbf{H}}_r}{{\mathbf{W}}_u}} \right) - {\overline \gamma  _{m}}\sigma _r^2 \ge 0,\\
\label{rank 1}&{{\mathbf{W}}_{m}}, {{\mathbf{W}}_{u}}\succeq 0, {{\mathbf{W}}_{m}}, {{\mathbf{W}}_{u}}\in {{\mathbb{H}}^N},\\
\label{rank}&{\rm{Rank}}\left( {{{\mathbf{W}}_{m}}} \right)=1, {\rm{Rank}}\left( {{{\mathbf{W}}_u}} \right) = 1,\\
\label{constraints beamformer 1}&\eqref{radar QoS},\eqref{power Communication},
\end{align}
\end{subequations}
\vspace{-0.6cm}

\noindent where \eqref{CT C QoS} and \eqref{CT T QoS} are arranged from \eqref{CT QoS 2}. Furthermore, we defined ${{\mathbf{H}}_c} \triangleq {{\mathbf{h}}_c}{\mathbf{h}}_c^H$, ${{\mathbf{H}}_r} \triangleq {{\mathbf{h}}_r}{\mathbf{h}}_r^H$,  and ${\overline \gamma  _{m}} = {2^{{{\overline R}_{m}}}} - 1$. Now, the non-convexity of the reformulated problem \eqref{beamformer based 1} only lies in the rank-one constraint \eqref{rank}. To tackle this obstacle, a popular technique is to use semidefinite relaxation (SDR)~\cite{protocol}. Explicitly, we firstly solve the problem by ignoring the rank-one constraint and then apply the Gaussian randomization method for constructing a rank-one solution, if the resultant solution is not of rank-one. However, considerable performance erosion may occur due to the reconstruction. On the other hand, it cannot be guaranteed that the reconstructed rank-one solution is still feasible in terms of satisfying all other constraints of the original problem (e.g., \eqref{radar QoS}, \eqref{CT C QoS}, and \eqref{CT T QoS}). As a remedy, a double-layer penalty-based iterative algorithm is proposed for gradually finding a near-optimal rank-one solution.\\ 
\indent To begin with, the non-convex rank-one constraint \eqref{rank} is equivalent to the following equality constraints:
\vspace{-0.2cm}
\begin{subequations}\label{DC rank}
\begin{align}\label{DC rank 1}&{\left\| {{{\mathbf{W}}_{m}}} \right\|_*} - {\left\| {{{\mathbf{W}}_{m}}} \right\|_2} = 0,\\
\label{DC rank 2}&{\left\| {{{\mathbf{W}}_{u}}} \right\|_*} - {\left\| {{{\mathbf{W}}_{u}}} \right\|_2} = 0,
\end{align}
\end{subequations}
\vspace{-0.6cm}

\noindent where ${\left\| \cdot  \right\|_ * }$ and ${\left\| \cdot \right\|_2} $ denote the nuclear norm and spectral norm of the matrix, respectively. Let us consider ${\mathbf{W}}_{m}$ as an example. It can be verified that, for any ${{\mathbf{W}}_{m}} \in {{\mathbb{H}}^N}$ and ${{\mathbf{W}}_{m}} \succeq 0$, the above equality constraint is only satisfied, when the matrix ${\mathbf{W}}_{m}$ is of rank-one. Otherwise, we always have ${\left\| {{{\mathbf{W}}_{m}}} \right\|_*} - {\left\| {{{\mathbf{W}}_{m}}} \right\|_2} > 0$.\\
\indent To solve problem \eqref{beamformer based 1}, we employ the penalty-based method of~\cite{penalty} by introducing the transformed equality constraints for ${\mathbf{W}}_{m}$ and ${\mathbf{W}}_{u}$ as a penalty term into the objective function of \eqref{beamformer based 1}, yielding the following optimization problem:
\vspace{-0.2cm}
\begin{subequations}\label{beamformer based 2}
\begin{align}
\mathop {\min }\limits_{{{\mathbf{W}}_m},{{\mathbf{W}}_u},{{\mathbf{R}}_1}}  &- {\rm{Tr}}\left( {{{\mathbf{H}}_c}{{\mathbf{W}}_u}} \right) + \frac{1}{\eta }\sum\limits_{i \in \left\{ {u,m} \right\}} {\left( {{{\left\| {{{\mathbf{W}}_i}} \right\|}_*} - {{\left\| {{{\mathbf{W}}_i}} \right\|}_2}} \right)} \\
\label{constraints beamformer 2}{\rm{s.t.}}\;\;&\eqref{radar QoS},\eqref{power Communication},\eqref{CT C QoS}-\eqref{rank 1},
\end{align}
\end{subequations}
\vspace{-0.6cm}

\noindent where $\eta  > 0$ is the penalty factor, which penalizes the violation of the equality constraints \eqref{DC rank 1} and \eqref{DC rank 2}, i.e., when ${\mathbf{W}}_{m}$ and ${\mathbf{W}}_{u}$ are not of rank-one. Since the maximization of ${R_{u}}$ is equivalent to maximizing the corresponding received signal strength of ${\rm{Tr}}\left( {{{\mathbf{H}}_c}{{\mathbf{W}}_u}} \right)$, we drop the $\log$ function in the objective function of \eqref{beamformer based 2} for simplicity. Despite relaxing the equality constraints in problem \eqref{beamformer based 2}, it may be readily verified that the solutions obtained will always satisfy the equality constraints (i.e., have rank-one matrices), when $\frac{1}{\eta } \to  + \infty $ ($\eta  \to 0$). However, if we firstly initialize $\eta $ with a sufficiently small value, the objective function's value of \eqref{beamformer based 2} tends to be dominated by the penalty term introduced, thus significantly degrading the efficiency of maximizing ${\rm{Tr}}\left( {{{\mathbf{H}}_c}{{\mathbf{W}}_u}} \right)$. To facilitate efficient optimization, we can initialize $\eta $ with a sufficiently large value to find a good starting point, and then gradually reduce $\eta $ to a sufficiently small value. As a result, feasible rank-one matrix solutions associated with a near-optimal performance can eventually be obtained. In the following, we will present the details of the double-layer penalty-based algorithm for solving problem \eqref{beamformer based 2}. In the inner layer, the optimization problem for a given $\eta $ is solved iteratively by employing successive convex approximation (SCA)~\cite{SCA} until convergence is reached. In the outer layer, the penalty factor, $\eta $, is gradually reduced from a sufficiently large value to a sufficiently small one.
\subsubsection{Inner Layer: Solving Problem \eqref{beamformer based 2} for A Given $\eta $} Note that for a given $\eta $, the non-convexity of \eqref{beamformer based 2} manifests itself in that the second term of each penalty term is non-convex, i.e., $ - {\left\| {{{\mathbf{W}}_{m}}} \right\|_2}$ and $ - {\left\| {{{\mathbf{W}}_u}} \right\|_2}$. However, they are concave functions with respect to both ${{{\mathbf{W}}_{m}}}$ and ${{{\mathbf{W}}_{u}}}$. By exploiting the first-order Taylor expansion, their upper bounds can be respectively expressed as follows:
\vspace{-0.2cm}
\begin{subequations}
\begin{align}\label{W1 uppder bound}&
\begin{gathered}
   - {\left\| {{{\mathbf{W}}_m}} \right\|_2} \le \overline {\mathbf{W}} _m^n \hfill \\
   \triangleq  - {\left\| {{\mathbf{W}}_m^n} \right\|_2} \!-\! {\rm{Tr}}\left[ {{{\mathbf{v}}_{\max }}\left( {{\mathbf{W}}_m^n} \right){{\mathbf{v}}_{{{\max }^H}}}\left( {{\mathbf{W}}_m^n} \right)\left( {{{\mathbf{W}}_m} - {\mathbf{W}}_m^n} \right)} \right], \hfill \\
\end{gathered} \\
\label{W2 uppder bound}&
\begin{gathered}
   - {\left\| {{{\mathbf{W}}_u}} \right\|_2} \le \overline {\mathbf{W}} _u^n \hfill \\
   \triangleq  - {\left\| {{\mathbf{W}}_u^n} \right\|_2} - {\rm{Tr}}\left[ {{{\mathbf{v}}_{\max }}\left( {{\mathbf{W}}_u^n} \right){{\mathbf{v}}_{{{\max }^H}}}\left( {{\mathbf{W}}_u^n} \right)\left( {{{\mathbf{W}}_u} - {\mathbf{W}}_u^n} \right)} \right], \hfill \\
\end{gathered}
\end{align}
\end{subequations}
\vspace{-0.5cm}

\noindent where ${{\mathbf{W}}_{m}^n}$ and ${{\mathbf{W}}_{u}^n}$ denote given points during the $n$th iteration of the SCA method, while ${{{\mathbf{v}}_{\max }}\left( {{\mathbf{W}}_{m}^n} \right)}$ and ${{{\mathbf{v}}_{\max }}\left( {{\mathbf{W}}_{u}^n} \right)}$ represent the eigenvector corresponding to the largest eigenvalue of ${{\mathbf{W}}_{m}^n}$ and ${{\mathbf{W}}_{u}^n}$, respectively.\\
\indent Accordingly, by exploiting the upper bounds obtained, problem \eqref{beamformer based 2} can be approximated by the following convex optimization problem:
\vspace{-0.2cm}
\begin{subequations}\label{beamformer based 3}
\begin{align}
\mathop {\min }\limits_{{{\mathbf{W}}_m},{{\mathbf{W}}_u},{{\mathbf{R}}_1}} &- {\rm{Tr}}\left( {{{\mathbf{H}}_c}{{\mathbf{W}}_u}} \right) + \frac{1}{\eta }\sum\limits_{i \in \left\{ {u,m} \right\}} {\left( {{{\left\| {{{\mathbf{W}}_i}} \right\|}_*} + \overline {\mathbf{W}} _i^n} \right)} \\
\label{constraints beamformer 3}{\rm{s.t.}}\;\;&\eqref{radar QoS},\eqref{power Communication},\eqref{CT C QoS}-\eqref{rank 1}.
\end{align}
\end{subequations}
\vspace{-0.6cm}

\noindent The above convex optimization problem can be efficiently solved by using existing standard convex problem solvers such as CVX~\cite{cvx}. Therefore, for a given $\eta$, problem \eqref{beamformer based 3} is iteratively solved until the fractional reduction of the objective function's value in \eqref{beamformer based 3} falls below the predefined threshold, $\epsilon_i$, when convergence is declared.
\subsubsection{Outer Layer: Reducing the Penalty Factor $\eta $}
In order to satisfy the equality constraints \eqref{DC rank 1} and \eqref{DC rank 2}, in the outer layer, we gradually update the value of $\eta$ towards a sufficiently small value as follows:
\vspace{-0.2cm}
\begin{align}\label{Update}
\eta  = \varepsilon \eta ,0 < \varepsilon  < 1,
\end{align}
\vspace{-0.6cm}

\noindent where $\varepsilon$ is a constant scaling factor, which has to be carefully selected for striking performance vs. complexity trade-off. For example, a larger $\varepsilon$ allows us to explore more potential candidate solutions, thus ultimately achieving a higher final performance. This, however, in turn requires more outer iterations hence imposing a higher complexity.
\subsubsection{Overall Algorithm and Complexity Analysis} Based on the above discussion, the proposed double-layer penalty-based procedure is summarized in \textbf{Algorithm 1}. The termination of the proposed algorithm depends on the violation of the equality constraints, which is expressed as follows:
\vspace{-0.2cm}
\begin{align}\label{termination}
\max \left\{ {{{\left\| {{{\mathbf{W}}_{m}}} \right\|}_*} - {{\left\| {{{\mathbf{W}}_{m}}} \right\|}_2},{{\left\| {{{\mathbf{W}}_u}} \right\|}_*} - {{\left\| {{{\mathbf{W}}_u}} \right\|}_2}} \right\} \le \epsilon_o ,
\end{align}
\vspace{-0.6cm}

\noindent where $\epsilon_o$ represents the maximum tolerable value. Upon reducing $\eta$, the equality constraints will finally be satisfied with the accuracy of $\epsilon_o$. According to ~\cite{SCA}, the proposed double-layer penalty-based algorithm is guaranteed to converge to a stationary point of the original problem \eqref{beamformer based 0}.\\
\indent The main complexity of \textbf{Algorithm 1} arises from iteratively solving problem \eqref{beamformer based 3}. Since problem \eqref{beamformer based 3} is a standard semidefinite program (SDP), the corresponding complexity is of the order of ${\mathcal{O}}\left( {2{N^{3.5}}} \right)$~\cite{Luo}. Therefore, the overall complexity of \textbf{Algorithm 1} is ${{\mathcal{O}}}\left( {{I_o^1}{I_i^1}\left( {2{N^{3.5}}} \right)} \right)$, where ${I_{i}^1}$ and ${I_{o}^1}$ denote the number of inner and outer iterations required for the convergence of \textbf{Algorithm 1}, respectively.
\vspace{-0.2cm}
\section{Numerical Results}
\begin{figure*}[htb!]
\centering
\begin{minipage}[t]{0.3\linewidth}
\includegraphics[width=2.4in]{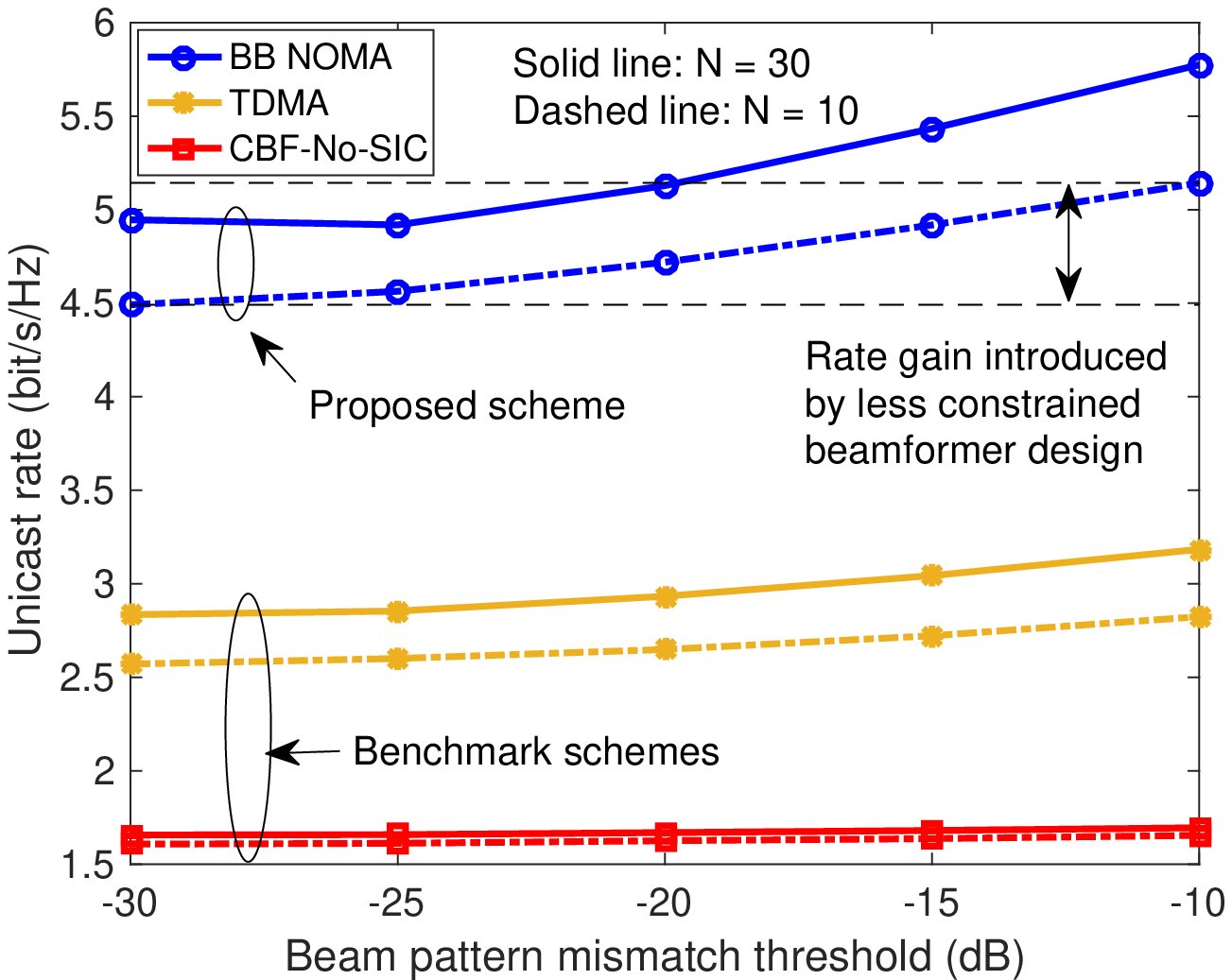}
\caption{The unicast rate versus ${\gamma _b}$.}
\label{Rvrb}
\end{minipage}
\quad
\begin{minipage}[t]{0.3\linewidth}
\includegraphics[width=2.4in]{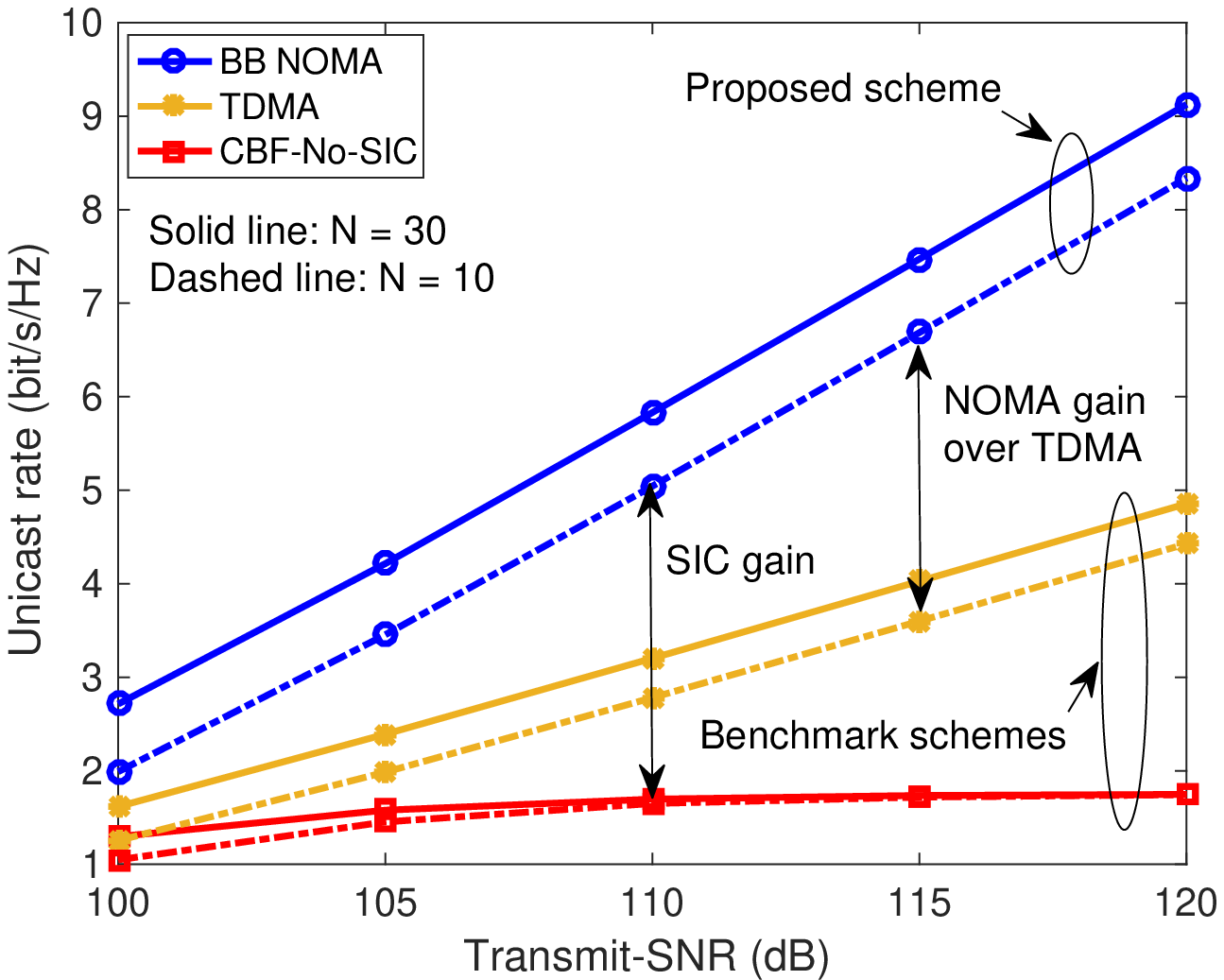}
\caption{The unicast rate versus ${\gamma _p}$.}
\label{RvP}
\end{minipage}
\quad
\begin{minipage}[t]{0.3\linewidth}
\includegraphics[width=2.4in]{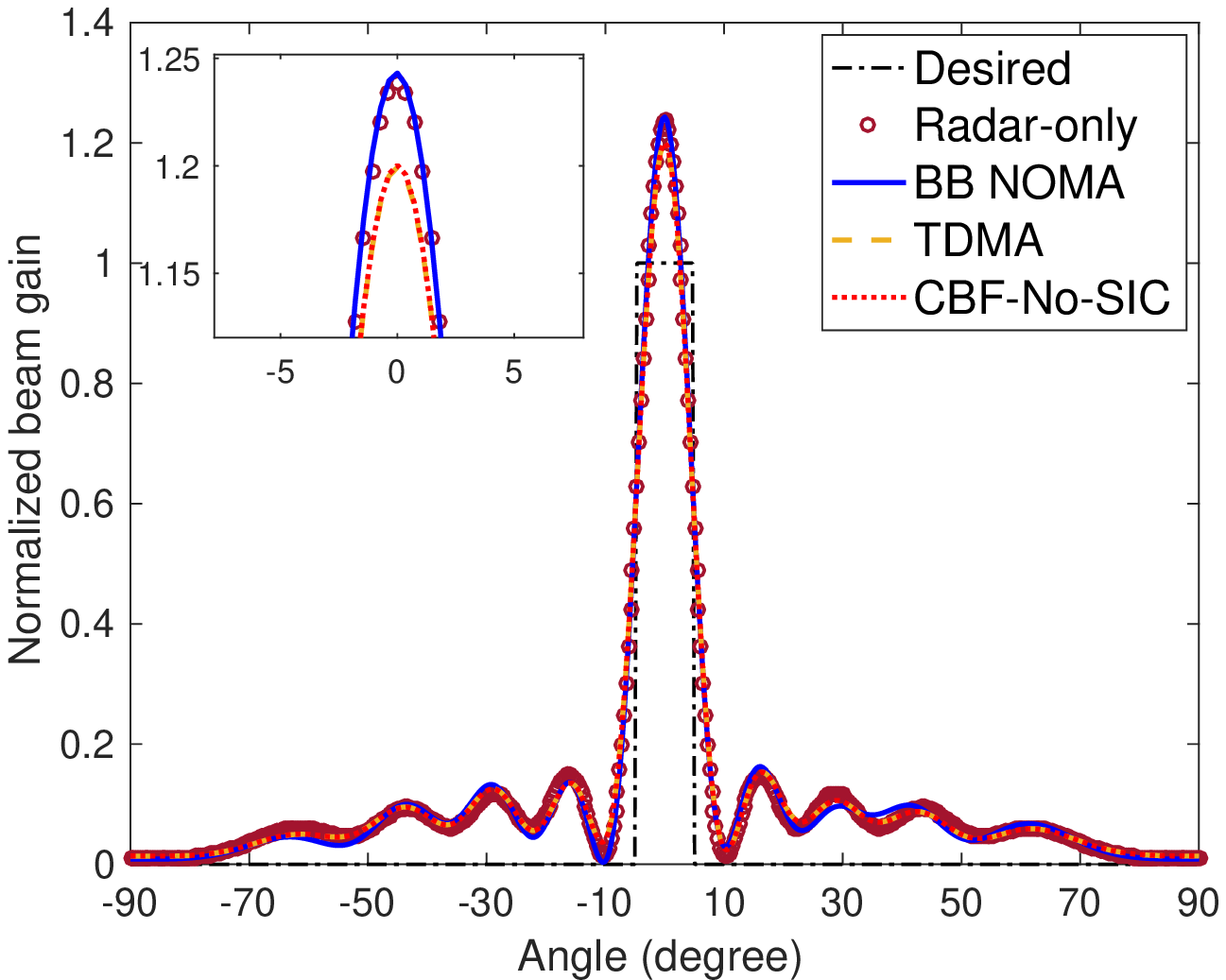}
\caption{The transmit beam pattern obtained by different schemes.}
\label{Beampattern_BB}
\end{minipage}
\end{figure*}

\vspace{-0.2cm}
In this section, we provide numerical results obtained by Monte Carlo simulations for characterizing the proposed NOMA-aided joint Rad-MU-Com frameworks. In particular, we assume that the DFRC BS employs a uniform linear array (ULA) with half-wavelength spacing between adjacent antennas. The channel between the BS and the R-user is assumed to have pure line-of-sight (LoS) associated with the path loss of ${L_R} = {L_0} + 20{\log _{10}}{d_R}$, while between the BS and C-user it is assumed to obey the Rayleigh channel model with the path loss of ${L_C} = {L_0} + 30{\log _{10}}{d_C}$~\cite{protocol,security}, where ${L_0}$ is the path loss at the reference distance $d=1$ meter (m), and ${d_R}$ and ${d_C}$ represents the distance from the BS to the R-user and to the C-user, respectively. The parameters adopted in simulations are set as follows: ${L_0}=40$ dB, ${d_R}=1000$ m, and ${d_C}=100$ m. The noise power in the receiver of users is assumed to be the same, which is set to ${\sigma ^2}=-100$ dBm. The transmit-signal-to-noise-ratio (SNR)\footnote{Using the transmit-SNR is unconventional, because it is given by the ratio of the transmit power and the receiver noise, which are quantities measured at different points. This quantity is however beneficial for our joint Rad-Com problem, where the optimum transmit power is assigned to each user for satisfying their individual rate requirements under the idealized simplifying assumption that they have perfect capacity-achieving receivers relying on powerful capacity-achieving channel codes. This is because the optimization problem of our specific system was formulated for maximizing the unicast performance at a given transmit power, while satisfying specific constraints imposed both on the multicast rate and on the radar beam pattern.} is considered in the simulations, which is given by ${\gamma _p} = \frac{{{P_{\max }}}}{{{\sigma ^2}}}$. The initial penalty factors of \textbf{Algorithm 1} is set to ${\eta _1} = {10^4}$, the convergence threshold of the inner layer is set to $\epsilon_i = 10^{-2}$, and the algorithm's termination threshold of the equality constraints is set to $\epsilon_o = 10^{-5}$. The numerical results were obtained by averaging over 200 channel realizations. \\
\begin{algorithm}[!t]\label{method1}
\caption{Proposed double-layer penalty-based algorithm for solving problem \eqref{beamformer based 0}}
\begin{algorithmic}[1]
\STATE {Initialize feasible points ${{\mathbf{W}}_{m}^0}$ and ${{\mathbf{W}}_{u}^0}$ as well as the penalty factor $\eta$.}
\STATE {\bf repeat: outer layer}
\STATE \quad Set iteration index $n=0$ for inner layer.
\STATE \quad {\bf repeat: inner layer}
\STATE \quad\quad For given ${{\mathbf{W}}_{m}^n}$ and ${{\mathbf{W}}_{u}^n}$, solve the convex problem \eqref{beamformer based 3} and the solutions obtained are denoted by ${{\mathbf{W}}_{m}^*}$ and ${{\mathbf{W}}_{m}^*}$.
\STATE \quad\quad ${\mathbf{W}}_{m}^{n + 1} = {\mathbf{W}}_{m}^*$, ${\mathbf{W}}_{u}^{n + 1} = {\mathbf{W}}_{u}^*$, and $n=n+1$.
\STATE \quad {\bf until} the fractional reduction of the objective function value falls below a predefined threshold $\epsilon_i >0$.
\STATE \quad ${\mathbf{W}}_{m}^0 = {\mathbf{W}}_{m}^*$, ${\mathbf{W}}_{u}^0 = {\mathbf{W}}_{u}^*$.
\STATE \quad Update $\eta  = \varepsilon \eta$.
\STATE {\bf until} the constraint violation falls below a maximum tolerable threshold $\epsilon_o >0$.
\end{algorithmic}
\end{algorithm}
\indent To obtain the actual true beam pattern, namely ${\mathbf{R}_0^*}$ of problem \eqref{ideal beam pattern design}, the desired beam pattern, ${P^*}\left( {{\theta _m}}\right)$, is defined as follows:
\vspace{-0.2cm}
\begin{align}\label{P}
{P^*}\left( {{\theta _m}} \right) = \left\{ \begin{gathered}
  1,\;\;{\theta _m} \in \left[ {{\overline \theta _k} - \frac{\Delta }{2},{\overline \theta _k} + \frac{\Delta }{2}} \right],\forall k \in {\mathcal{K}} \hfill \\
  0,\;\;{\rm{otherwise}}, \hfill \\
\end{gathered}  \right.
\end{align}
\vspace{-0.3cm}

\noindent where $\left\{ {{\overline \theta _k},\forall k \in {\mathcal{K}}} \right\}$ denotes the actual true angles to be detected, which are determined by the location of R-users, and $\Delta $ denotes the width of the desired beam, which is set to ${10^ \circ }$ in the simulation. For performance comparison, we consider two benchmark schemes, which have been discussed in \textbf{Remark 2}.\\
\indent In Fig. \ref{Rvrb}, we investigate the unicast rate, $R_u$, achieved versus the maximum tolerable beam pattern mismatch, ${{\overline \gamma}_{b}}$. We set ${\gamma _p}=110$ dB (i.e., ${P_{\max}}=10$ dBm) and ${\overline R}_{m}=0.5$ bit/s/Hz. As seen in Fig. \ref{Rvrb}, the unicast rate obtained by all schemes increases as ${{\overline \gamma}_{b}}$ increases. This is indeed expected, since lager ${{\overline \gamma}_{b}}$ values impose looser constraints on the BF design, which provides higher DoFs, hence enhancing the unicast performance. Moreover, a higher $N$ leads to a higher unicast rate due to the enhanced array gain and spatial DoFs. By comparing the three Rad-MU-Com schemes presented, it may be observed that the proposed BB NOMA+Rad-MU-Com scheme achieves the best performance. This is because on the one hand, employing SIC in NOMA mitigates the inter-signal interference compared to the CBF-No-SIC+Rad-MU-Com scheme, thus improving the unicast performance achieved at the C-user. On the other hand, the power-domain resource sharing and the employment of two different BFs allows NOMA to achieve a higher performance than the TDMA+Rad-MU-Com scheme. Moreover, despite employing a single BF, the TDMA+Rad-MU-Com scheme can deliver both types of messages in an interference-free manner, while carrying out radar detection. Therefore, the TDMA+Rad-MU-Com scheme outperforms the CBF-No-SIC+Rad-MU-Com scheme, whose performance is significantly degraded by the inter-signal interference. It can also be observed that the performance gain obtained by NOMA is more noticeable when ${{\overline \gamma}_{b}}$ increases. The above results verify the efficiency of the proposed BB NOMA+Rad-MU-Com framework.\\
\indent In Fig. \ref{RvP}, we present the unicast rate, $R_u$, achieved versus the transmit-SNR, ${\gamma _p}$. We set ${\overline R}_{m}=0.5$ bit/s/Hz and ${{\overline \gamma}_{b}}=-10$ dB. It can be observed that the unicast rate of all schemes increases upon increasing ${\gamma _p}$. However, in contrast to both NOMA and TDMA, the rate enhancement of CBF-No-SIC attained upon increasing ${\gamma _p}$ becomes negligible and the unicast rate is seen to be bounded by a certain value. This is because when the inter-signal interference is not mitigated, CBF-No-SIC becomes interference-limited, when the transmit power is high. Moreover, it can also be seen from Fig. \ref{RvP} that the rate enhancement attained by NOMA upon increasing ${\gamma _p}$ is more significant than for TDMA, since NOMA benefits from a flexible resource allocation scheme.\\
\indent In Fig. \ref{Beampattern_BB}, we plot the transmit beam pattern obtained by the three schemes for one random channel realization. We set $N=10$, ${\gamma _p}=105$ dB, ${\overline R}_{m}=0.5$ bit/s/Hz, and ${{\overline \gamma}_{b}}=-20$ dB. In particular, the desired beam pattern is obtained according to \eqref{P} and the beam pattern of the radar-only system is obtained by solving problem \eqref{ideal beam pattern design}. As illustrated in Fig. \ref{Beampattern_BB}, the beam pattern obtained by NOMA closely approaches that of the radar-only system, while the beam pattern mismatch of TDMA and CBF-No-SIC becomes more noticeable. Observe in Figs. \ref{Rvrb} and \ref{RvP} that given the same accuracy requirement of the radar beam pattern, the proposed BB NOMA+Rad-MU-Com scheme achieves higher communication performance than the other benchmark schemes. The above results also confirm the effectiveness of the proposed BB NOMA+Rad-MU-Com framework.
\vspace{-0.2cm}
\section{Conclusions}
\vspace{-0.1cm}
A novel BB NOMA-aided joint Rad-MU-Com framework has been proposed, where a MIMO DFRC BS transmits superimposed multicast and unicast messages to the R- and C-users, while detecting the R-user target. Specifically, a BF optimization problem was formulated for enhancing the unicast performance, while satisfying both the multicast rate and the radar beam pattern requirements. To solve this problem, a penalty-based iterative algorithm was developed to find a near-optimal solution. The numerical results revealed that a higher unicast performance can be achieved by the proposed BB NOMA-aided joint Rad-MU-Com schemes than by the TDMA based scheme and the scheme without SIC.

\vspace{-0.2cm}
\bibliographystyle{IEEEtran}
\bibliography{mybib}

\end{document}